# Giant and Flexible Toroidal Circular Dichroism from Planar Chiral Metasurface

**Shijie Kang[1,2], Haitao Li[1], Jiayu Fan[1], Jiusi Yu[1], Boyang Qu[1,2], Peng Chen[1,3], Xiaoxiao Wu[1,2,*)]**

[1]*Modern Matter Laboratory and Advanced Materials Thrust, The Hong Kong University of Science and Technology (Guangzhou), Nansha, Guangzhou 511400, China*

[2]*Low Altitude Systems and Economy Research Institute, The Hong Kong University of Science and Technology (Guangzhou), Nansha, Guangzhou 511400, Guangdong, China*

[3]*Quantum Science center of Guangdong-Hong Kong-Macao Greater Bay Area, Shenzhen-Hong Kong International Science and Technology Park, Futian, Shenzhen 518000, Guangdong, China*

**xiaoxiaowu@hkust-gz.edu.cn*

**Abstract:** Chirality, a fundamental concept describing an object cannot superpose with its mirror image, is crucial in optics and photonics and leads to various exotic phenomena, such as circular dichroism, and optical activity. Recent findings reveal that, besides electric and magnetic dipoles, toroidal dipoles, an elusive part of dynamic multipoles, can also contribute significantly to chirality. However, as toroidal dipoles are typically represented by solenoidal currents circulating on a three-dimensional (3D) torus, toroidal circular dichroism is usually observed in 3D intricate microstructures. Facing corresponding challenges in fabrication, integration and application, it is generally difficult to employ toroidal circular dichroism in compact metasurfaces for flexible modulation of chiral interactions between electromagnetic waves and matter. To overcome these stringent challenges, we propose and experimentally demonstrate the giant toroidal circular dichroism in a bilayer metasurface that is comprised of only planar layers, effectively bypassing various restrictions imposed by 3D microstructures.

With the introduction of a displacement, or bilayer offset, between the opposite layers, we experimentally achieve giant chiral responses with the intrinsic circular dichroism (CD) reaching 0.69 in measurements, and the CD can be quantitatively manipulated in a simple manner. The giant intrinsic chirality primarily originates from distinct excitations of in-plane toroidal dipole moments under circular polarized incidences, and the toroidal chiral response is quantitatively controlled by the bilayer offset. Therefore, our work provides a straightforward and versatile approach for development of giant and flexible intrinsic chirality through toroidal dipoles with inherently planar layers, important for applications in communications, sensing, and chiroptical devices.

## 1. Introduction

Chiral metamaterials are ensembles of microstructures which satisfy chirality, meaning that they cannot be superimposed on their mirror images [1-3]. When chiral metamaterials interact with circularly polarized waves, they will exhibit much stronger circular dichroism (CD) compared with natural materials, where the absorption or transmission can vary significantly for left and right circular polarizations (LCP and RCP) [4-6]. To unlock the circular polarization as a new degree of freedom [7], they hold particular importance in both theoretical research and practical applications of modern optics and photonics. For example, the phenomenon of CD and optical activity have been explored in the last decade using elaborately designed three-dimensional (3D) microstructures [8, 9], such as helix [10] and multi-layered [11, 12] constructions which can show a strong chiral response. However, these bulky 3D designs impose considerable fabrication challenges as they often exceed the capabilities of traditional fabrication techniques due to their complex geometries and spatial dimensions constraints [13]. Very recently, a variety of chiral metasurfaces, which are thin-layer [14-17] ensembles of microstructures arranged in a two-dimensional (2D) manner [18-20], have been designed to exhibit strong CD ($\geq 0.5$) responses. Working from GHz microwaves to the visible light, the common approaches to induce chirality for the metasurfaces include out-of-plane symmetry breaking [21, 22], in-plane symmetry breaking [23], and spatial inversion symmetry breaking [24]. Unfortunately, these symmetry-breaking perturbations generally depend on 3D microstructures, and the corresponding metasurfaces are inherently non-planar. As a direct result, most such metasurfaces are quite thick in the out-of-plane direction, and their chirality strongly depends on the thickness, posing stark difficulties for fabrication, integration, and applications [25].

It is discovered that toroidal dipoles, an elusive part of dynamic multipoles and long thought as a localized excitation, can also contribute significantly to the intrinsic chirality, leading to so-called "toroidal circular dichroism" (toroidal CD). However, strong toroidal dipoles are usually excited by incidences on 3D microstructures, especially under the case of normal incidences which correspond to intrinsic chirality. Thus, it is generally believed that toroidal CD can hardly emerge from deep subwavelength (<1/10 wavelength) metasurfaces which can be considered truly planar for the incident EM wave. Hence, toroidal CD, despite being an exotic phenomenon capable of generating strong intrinsic chirality, faces substantial hurdles in its applications to flexibly modulate the chiral interactions between the electromagnetic (EM) waves and matter.

Inspired by recent development in multi-layer metasurfaces [26-30], we propose and experimentally demonstrate the giant toroidal CD in a bilayer metasurface to overcome these significant hurdles and barriers. The metasurface is comprised of only thin planar layers, effectively bypassing various restrictions imposed by 3D microstructures. In fact, our study proposes an additional degree of freedom — the bilayer offset — as a simple and flexible asymmetry parameter to modulate the bilayer metasurface, evoking a strong intrinsic CD response. The findings, both in simulations and experiments, demonstrate that the CD response of the metasurface depends on this simple asymmetry parameter and peaks when it equals to half of the lattice constant. Moreover, the multipole scattering cross sections retrieved from full-wave simulations reveal that in-plane toroidal dipole moments plays the major role in the giant chiral response, confirming the planar metasurface exhibiting strong toroidal CD [31]. This fact is a departure from the common expectation that toroidal dipoles can only lead to strong intrinsic chiral responses in complex 3D microstructures. In this way, we also provide a general and straightforward scheme for designing future planar chiral metasurfaces that should be much easier for on-chip integration, which are important for next-generation chiral devices in applications of telecommunications, sensing, spectroscopy, and beyond.

## 2. Results

An illustration of our designed bilayer metasurface, inspired by the famous split rings [32, 33], but with necessary symmetries broken to realize chiral response, is shown in Fig. 1(a). It conceptually demonstrates the toroidal CD phenomenon of the planar metasurface under normal incidence of circular-polarized EM waves, that is, how the planar metasurface exhibits strikingly different transmission responses for LCP and RCP waves. As also artistically depicted in Fig. 1(a), this intrinsic CD is mainly induced by in-plane toroidal dipoles, which will be further discussed later. To clarify its design principle, we start from an achiral metasurface unit cell shown in Fig. 1(b), comprising copper split rings of the identical shape deposited on both sides of the F4B substrate. To induce desired chiral response, we need to introduce asymmetry to this initial unit cell with a series of geometry operations, as shown in Fig. 1(c). First, a 180° rotation operation is applied to the upper metallic layer in each unit cell, and then the upper layer is further shifted along the $y$ direction with a distance of $d$. Without loss of generality, we assume the geometric parameters for the unit cell, the lattice constant $a$ = 10 mm, the width of the split and the ring $w$ = 0.3 mm, the thickness of the substrate $t_{sub}$ = 2 mm, and the side length of the square $b$ = 7.2 mm. Our proposed microstructure with bilayer offset $d$ provides a simple way to break the spatial inversion symmetry and mirror symmetry of the metasurface [34], making it possible to achieve significant chiral response with only planar layers. In the following study, we will show that with different shifting movements represented by different $d$ values, also referred to below as called bilayer offset, the bilayer metasurface can be flexibly tuned from no chirality to the giant chirality case.

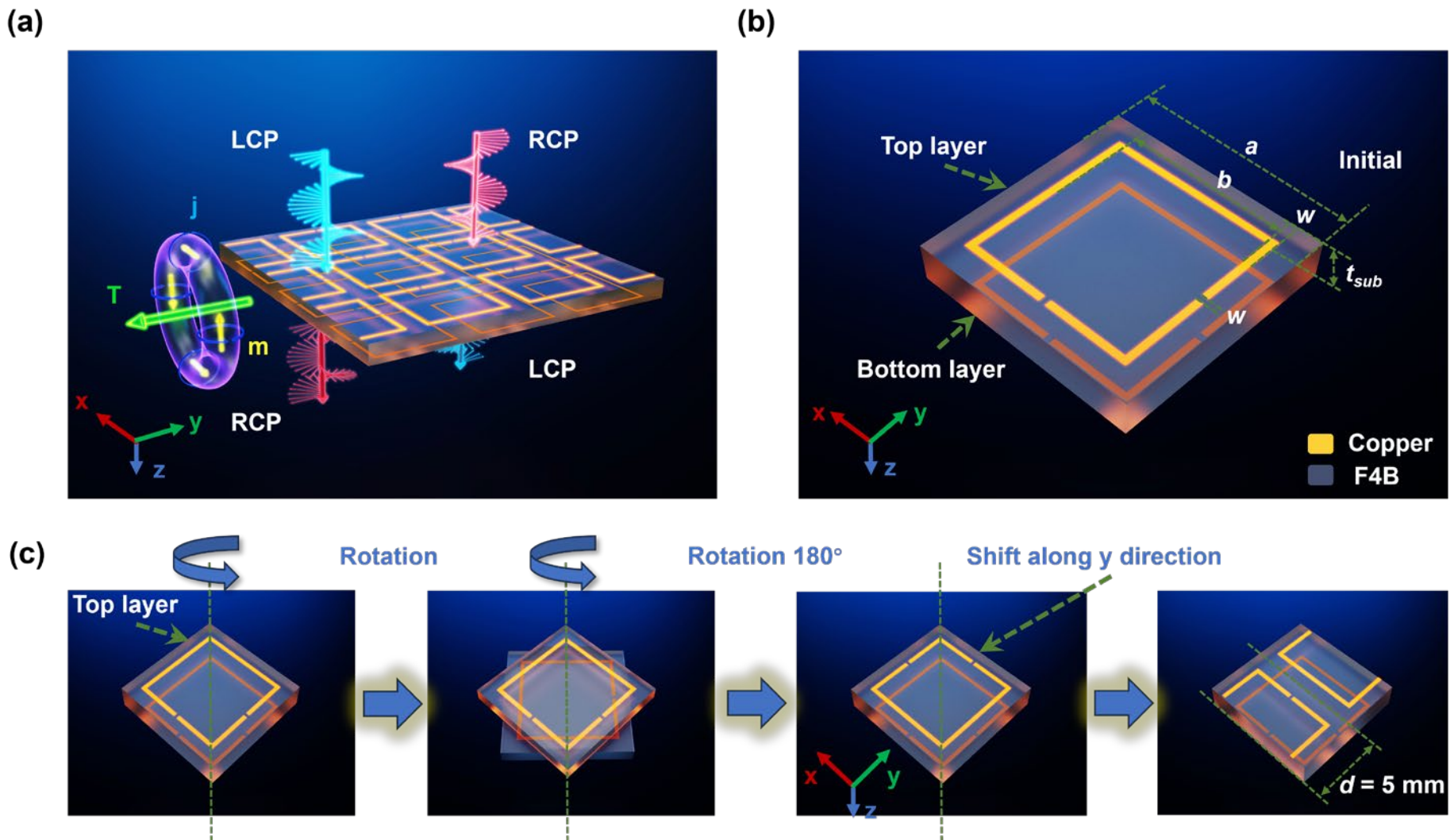

**Fig. 1. Illustration of the proposed metasurface and its construction.** (a) Conceptual depiction of the intrinsically chiral metasurface under the normal incidences of left circular polarized (LCP) and right circular polarized (RCP) EM waves. An in-plane toroidal dipole moment **T** (green) is artistically

represented by poloidal currents **j** (bule) circulating on a torus to emphasize the toroidal circular dichroism of the metasurface. (b) Schematic of the initial achiral unit cell structure and its geometrical parameters. Square split rings are periodically deposited on the top and bottom surfaces of the structure consisting of metallic patterns (yellow) on a dielectric F4B substrate (blue gray). (c) Rotation and shifting operations are applied to the top surface of the initial unit cell to break the symmetries to induce chirality.

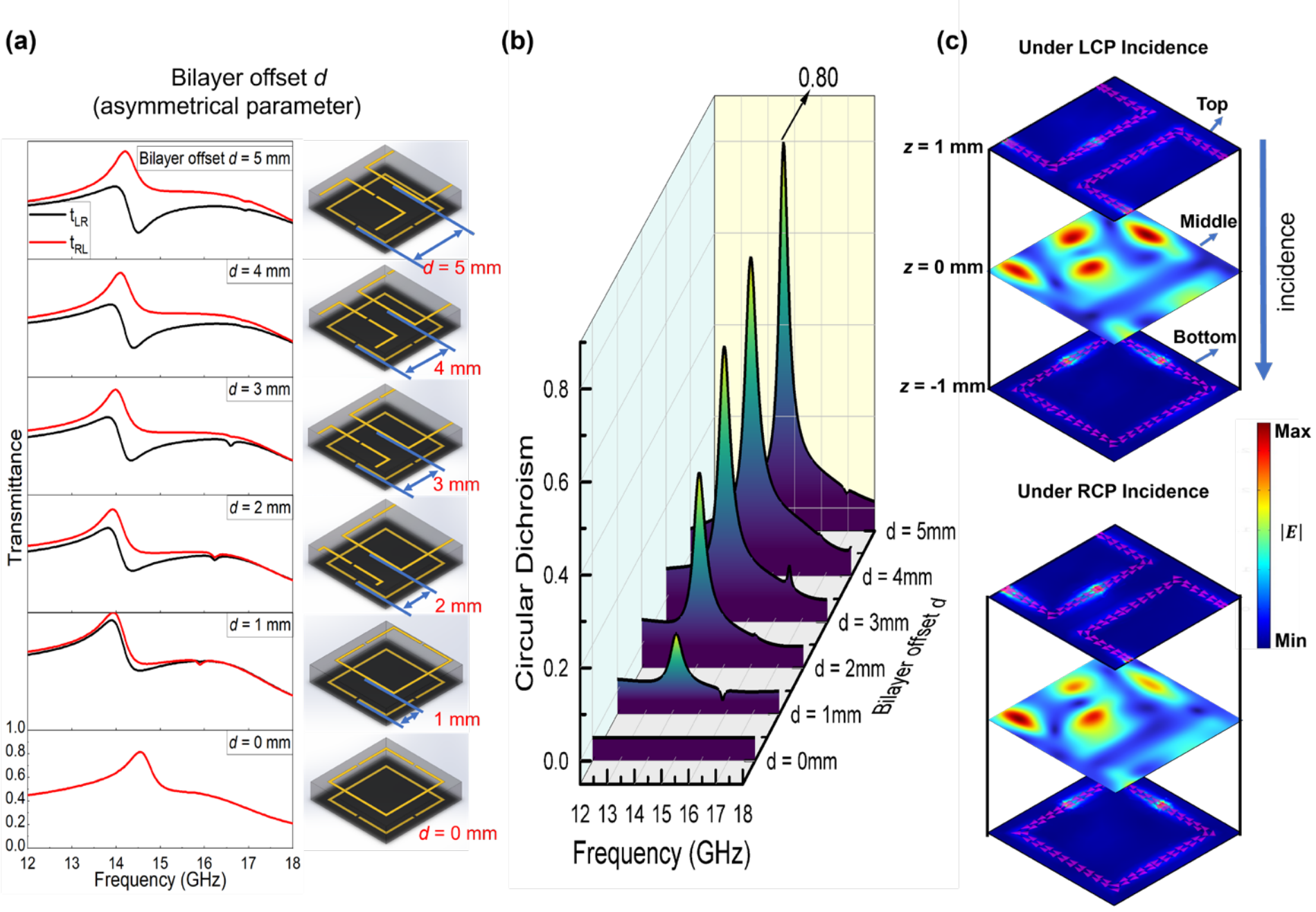

**Fig. 2. Different chirality of the metasurface under varying values of the bilayer offset *d*.** (a) Cross-polarization transmission spectra ($t_{LR}$ and $t_{RL}$) under LCP and RCP wave incidences for different bilayer offset *d*. (b) Retrieved CD spectra for different bilayer offset *d* and the maximum CD value can reach 0.80 when *d* value is set to be half of the lattice constant *a*. (c) Three slices of the electric field amplitude (|E|) together with the surface current distributions (magenta arrows) at the frequency of the maximum CD for bilayer offset *d* = 5 mm under LCP and RCP wave incidences, respectively.

In fact, as shown in Fig. 2(a), we simulate the circular polarization transmissions of the metasurfaces with different bilayer offset $d$ using COMSOL Multiphysics, a commercial finite-element method software package. As our metasurface is periodic, the different values of bilayer offset $d$ can be limited between 0 (no shifting) to the lattice constant $a$ = 10 mm. For clear visualization, we only demonstrate the cross-polarization transmission spectra $t_{LR}$ and $t_{RL}$ from $d$ = 0 mm to 5 mm, half of the lattice constant $a$ (see Supplement 1 Figure S1 for remaining values of $d$ from 6 mm to 10 mm). The calculated co-polarized transmission spectra $t_{LL}$ and $t_{RR}$ for different bilayer offset $d$ are almost identical (see Supplement 1 Figure S2). From Fig. 2(a) we can clearly see that the cross-polarized transmission spectra $t_{LR}$ and $t_{RL}$ overlap when there is no bilayer offset ($d$ = 0 mm). Meanwhile, as $d$ value increases, the difference between $t_{LR}$ and $t_{RL}$ becomes larger and larger and is maximized when the value of d reaches half of the lattice constant $a$, that is, $d = a/2 = 5$ mm. Moreover, the profound differences between cross-polarized transmission coefficients ($t_{LR}$ and $t_{RL}$) suggest that there can be a giant chiral response of our proposed metasurface. With the definition of the CD value as [35, 36]

$$CD = \frac{t_{LCP} - t_{RCP}}{t_{LCP} + t_{RCP}} = \frac{\left(\left|t_{LL}\right|^2 + \left|t_{RL}\right|^2\right) - \left(\left|t_{RR}\right|^2 + \left|t_{LR}\right|^2\right)}{\left(\left|t_{LL}\right|^2 + \left|t_{RL}\right|^2\right) + \left(\left|t_{RR}\right|^2 + \left|t_{LR}\right|^2\right)} , \qquad (1)$$

where $t_{ij}$ ($i, j = L, R$, and $L$ stands for LCP while $R$ stands for RCP) represents the transmission coefficients under circular polarizations, and $t_{\mathrm{LCP}} = |t_{LL}|^2 + |t_{RL}|^2$ ($t_{\mathrm{RCP}} = |t_{RR}|^2 + |t_{LR}|^2$) represents the total transmission for LCP (RCP) incidence. The CD spectra of the metasurfaces with $d$ values from 0 to 5 mm are plotted in Fig. 2(b), and we can observe a giant CD value (CD = 0.80) at the resonance frequency 14.3 GHz for the metasurface when $d$ = 5 mm. The CD spectra of the remaining $d$ values (6 mm to 10 mm) are enclosed in the Supplement 1 (Figure S3), and generally for the CD spectra, the cases of $d$ and $a-d$ are almost identical (such as $d$ = 6 mm and $d$ = 4 mm). Therefore, we can see that the CD spectra vary with the value of the bilayer offset $d$, from no chiral response (CD = 0) when $d$ = 0 mm to the maximum chiral response when $d$ = 5 mm, and back to no chiral response when $d$ = 10 mm. Therefore, we can identify the bilayer offset $d$ as the asymmetry parameter which control the chirality of the bilayer metasurface [36]. Moreover, we have also calculated the impact of other geometric parameters, such as the side length $b$, on the chiral response of the metasurface. It is found that they mainly shift the resonance frequency of the metasurface, while only slightly changing peak values of the CD spectra (see Figure S8 in Supplement 1 for detailed discussion).

To further investigate the physical origin of the giant chiral response generated by the bilayer planar metasurface, we performed simulations to calculate the spatial distributions of the electric field $\mathbf{E}$ and corresponding induced surface current density $\mathbf{J}_s$ in the planar chiral metasurface at the resonant frequency of 14.3 GHz, with a bilayer offset of $d$ = 5 mm. In Fig. 2(c), we illustrate the distribution of the electric field amplitude $|\mathbf{E}|$ across three slices (top, middle, and bottom surfaces of the unit cell) of the metasurface under LCP and RCP wave incidences, respectively, together with the induced surface current densities. It is evident that under LCP and RCP wave incidences, the electric field $|\mathbf{E}|$ is both concentrated at the $z$ = 0 mm slice due to the resonance; however, the electric field is enhanced around different spots. More importantly, the surface current density $\mathbf{J}_s$ indicated by magenta arrows on the split rings under the LCP and RCP incidences exhibit extremely different circulations. These differences indicate potentially different excitations of certain multipoles in the metasurface structure under LCP and RCP incidences. In fact, it is the significantly different excitations of toroidal and magnetic dipole moments that contribute mainly to the giant intrinsic chirality discovered here, which will be discussed later in the work.

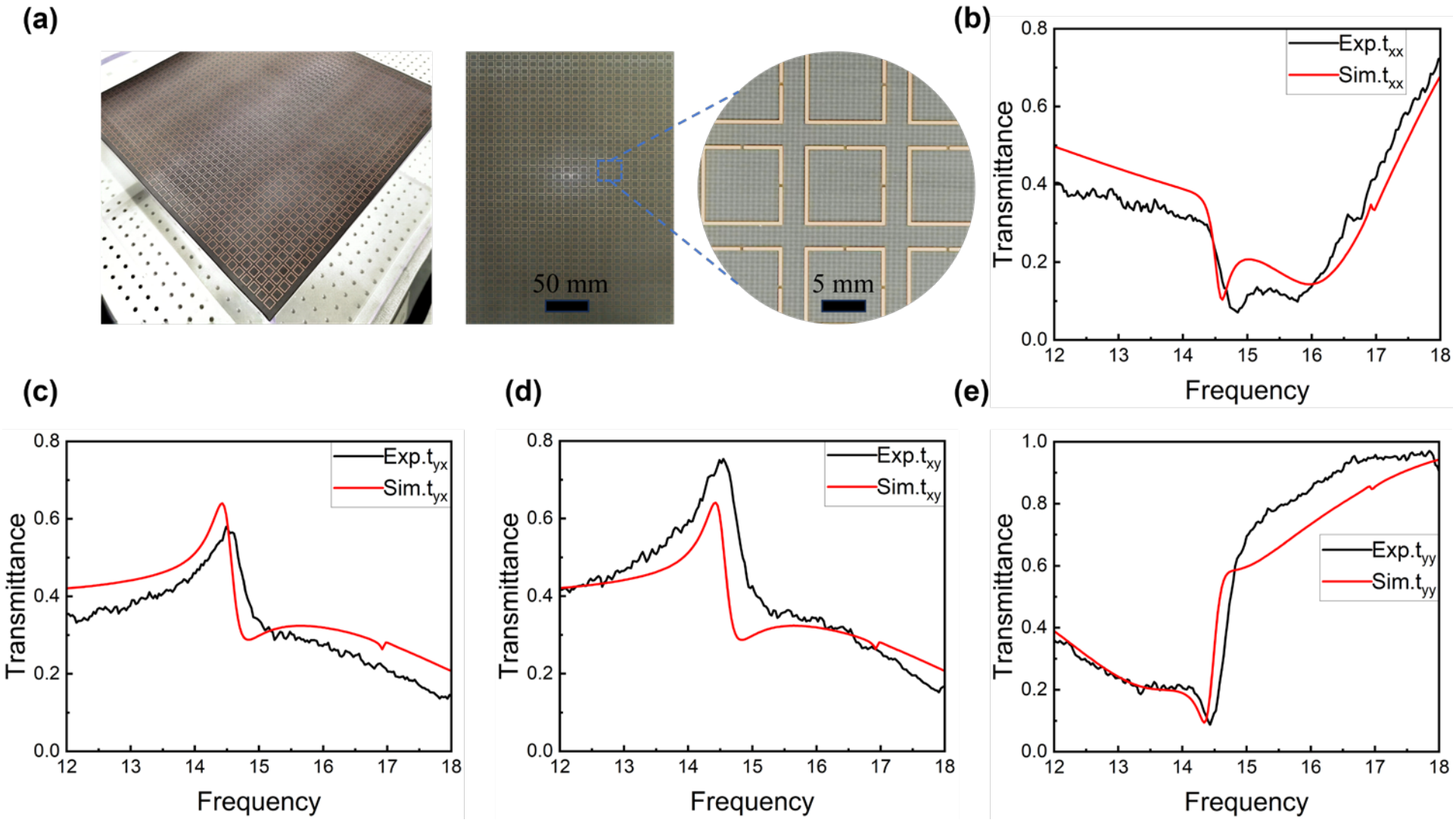

**Fig. 3. The fabricated sample and linear transmission coefficients spectra.** (a) The photograph of the fabricated sample including isometric projection, top view and inset of enlarged details of the unit cell. (b-e) Measured (black solid line) and simulated (red solid line) transmission coefficients spectra ($t_{yy}$, $t_{xx}$, $t_{yx}$ and $t_{xy}$) for the bilayer offset $d$ = 5 mm with linear polarizations.

To experimentally verify our results (see Supplement 1 Figure S4 and Note 1 for details of experimental setup), samples with bilayer offset $d$ = 5 mm corresponding to the maximum CD effect are fabricated according to our designs by the printed circuit board (PCB) techniques. The photographs of the fabricated samples including isometric view, top view and enlarged view are shown in Fig. 3(a). To investigate the impact of the metasurface on polarizations, the x- and y-polarized transmitted EM wave is denoted as $[E_x^t, E_y^t]^T$ [37], while the incident EM wave impinging on the metasurface is denoted as $[E_x^i, E_y^i]^T$ With the Jones matrix [38, 39], we have

$$\begin{pmatrix} E_x^t \\ E_y^t \end{pmatrix} = T \begin{pmatrix} E_x^i \\ E_y^i \end{pmatrix} = \begin{pmatrix} t_{xx} & t_{xy} \\ t_{yx} & t_{yy} \end{pmatrix} \begin{pmatrix} E_x^i \\ E_y^i \end{pmatrix} , \tag{2}$$

where $t_{xx}$ and $t_{yy}$ are the corresponding linear co-polarization transmission coefficients and $t_{xy}$ and $t_{yx}$ are the linear cross-polarization transmission coefficients. To obtain the linear transmission coefficients through far-field measurements, the sample under test is placed between two linear-polarized horn antennas which is connected to a vector network analyzer (VNA, Ceyear 3674H, China). Both the measured and simulated transmission coefficients in linear polarizations are plotted in Fig. 3(b-e) for comparison. All the measured results are essentially in agreement with the ones from the full-wave simulations (see Supplement 1 Note 2 for details of simulation settings using COMSOL).

Based on the outcomes of the linear polarized transmissions, we then convert them to circular-polarized transmission coefficients according to the following relationship [40]

$$\begin{pmatrix} t_{RR} & t_{RL} \\ t_{LR} & t_{LL} \end{pmatrix} = \frac{1}{2}\begin{pmatrix} \mathrm{t}_{xx} + t_{yy} + i\left(t_{xy} - t_{yx}\right) & t_{xx} - t_{yy} - i\left(t_{xy} + t_{yx}\right) \\ t_{xx} - t_{yy} + i\left(t_{xy} + t_{yx}\right) & t_{xx} + t_{yy} - i\left(t_{xy} - t_{yx}\right) \end{pmatrix}, \tag{3}$$

Using Eq. (3), we calculate the circular polarized transmission spectra from both experimental measurements and full-wave simulations, as is shown in Fig. 4 (a-d). It can be seen they are in good agreement. We then sum the transmissions for LCP and RCP incidence and plotted them in Fig. 4 (e), confirming that there is significant chiral response for cross-polarization transmissions [41]. However, there are some discrepancies between experimental and simulated transmission measurements, particularly the oscillatory behavior and closely aligned peaks near 14.5 GHz, arising from several possible factors. These factors include diffraction and edge scattering due to the finite size of the metasurface, misalignment or polarization mismatches of the horn antennas, fabrication tolerances and material imperfections, and limitations in the simulation models regarding boundary conditions and material properties. Together, these factors could cause deviations that are not fully captured in the simulations. Moreover, the CD spectrum from measurements is calculated according to Eq. (1) and displayed in Fig. 4 (f), and indeed shows the giant CD response (peak value 0.80). The thickness of our metasurface is 2mm which is smaller than 1/10 of the working wavelength, and can be further decreased, for example, to around 1/40, while it still exhibits strong (CD ≥ 0.5 around resonance) chiral responses (see Supplement 1 Figure S5 for calculated CD values for the metasurfaces with other ultra-thin thicknesses). Therefore, the significant chirality of our metasurface is robust with respect to the substrate thickness. We also note that all simulations and experiments correspond to the normal incidence case, thus the chirality of our ultra-thin metasurface is intrinsic chirality [42-44]. Moreover, the chirality of metasurface will not be degraded when there is moderate dielectric loss in the substrate (see Supplement 1 Figure S6), hence our design principle could also be applied to higher frequency regions, such as terahertz (THz), where the dielectric loss becomes non-negligible.

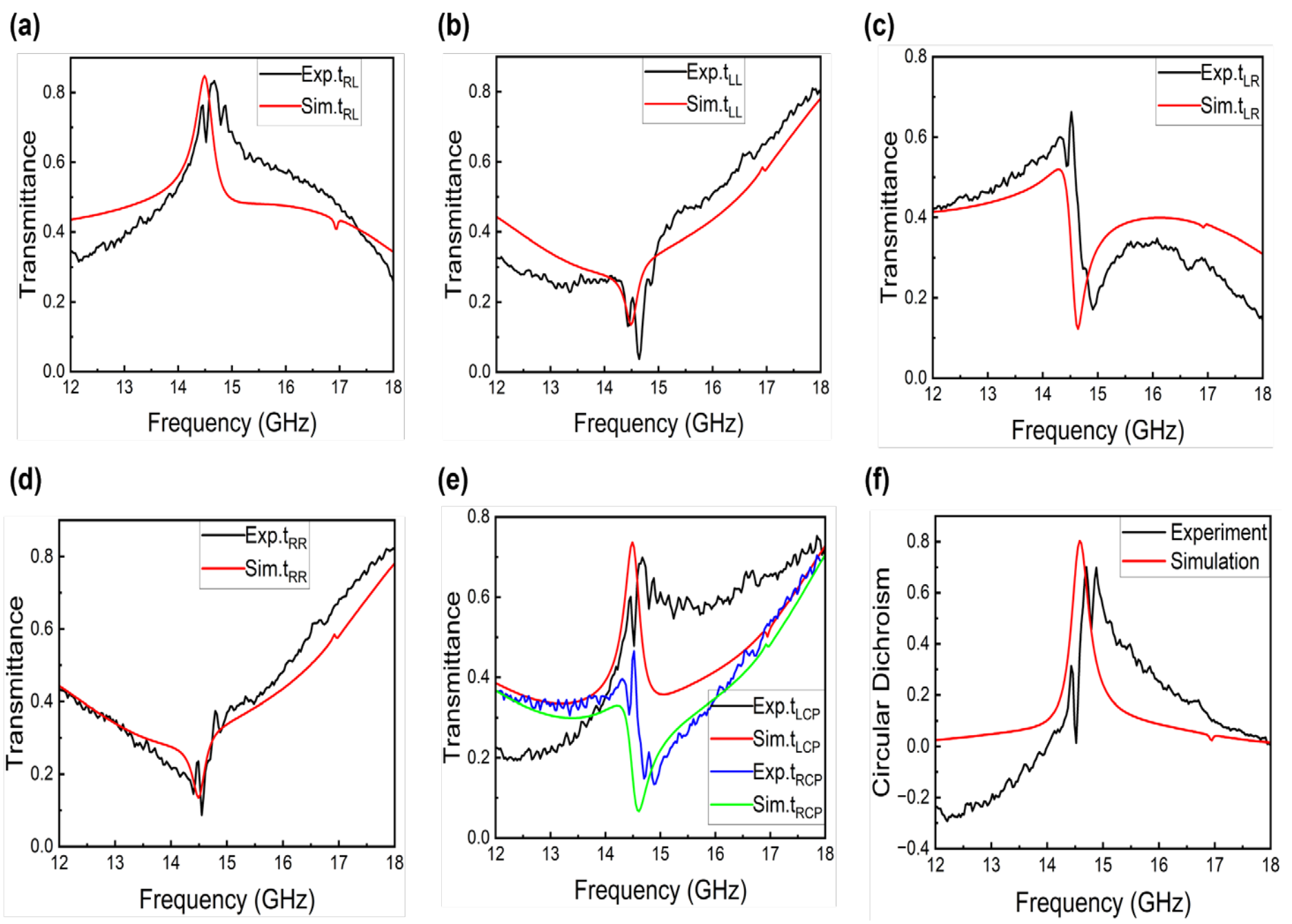

**Fig. 4. The comparison of circular transmission spectra from experiments and simulations.** (a-d) Simulated and experimental circular-polarized transmission spectra including cross-polarized ($t_{RL}$ and $t_{LR}$) and co-polarized ($t_{LL}$ and $t_{RR}$) transmission coefficients. (e) Simulated and experimental $t_{LCP}$ and $t_{RCP}$

spectra, summing the transmitted polarizations. (f) Simulated and measured circular dichroism (CD) spectra with the bilayer offset $d$ = 5 mm.

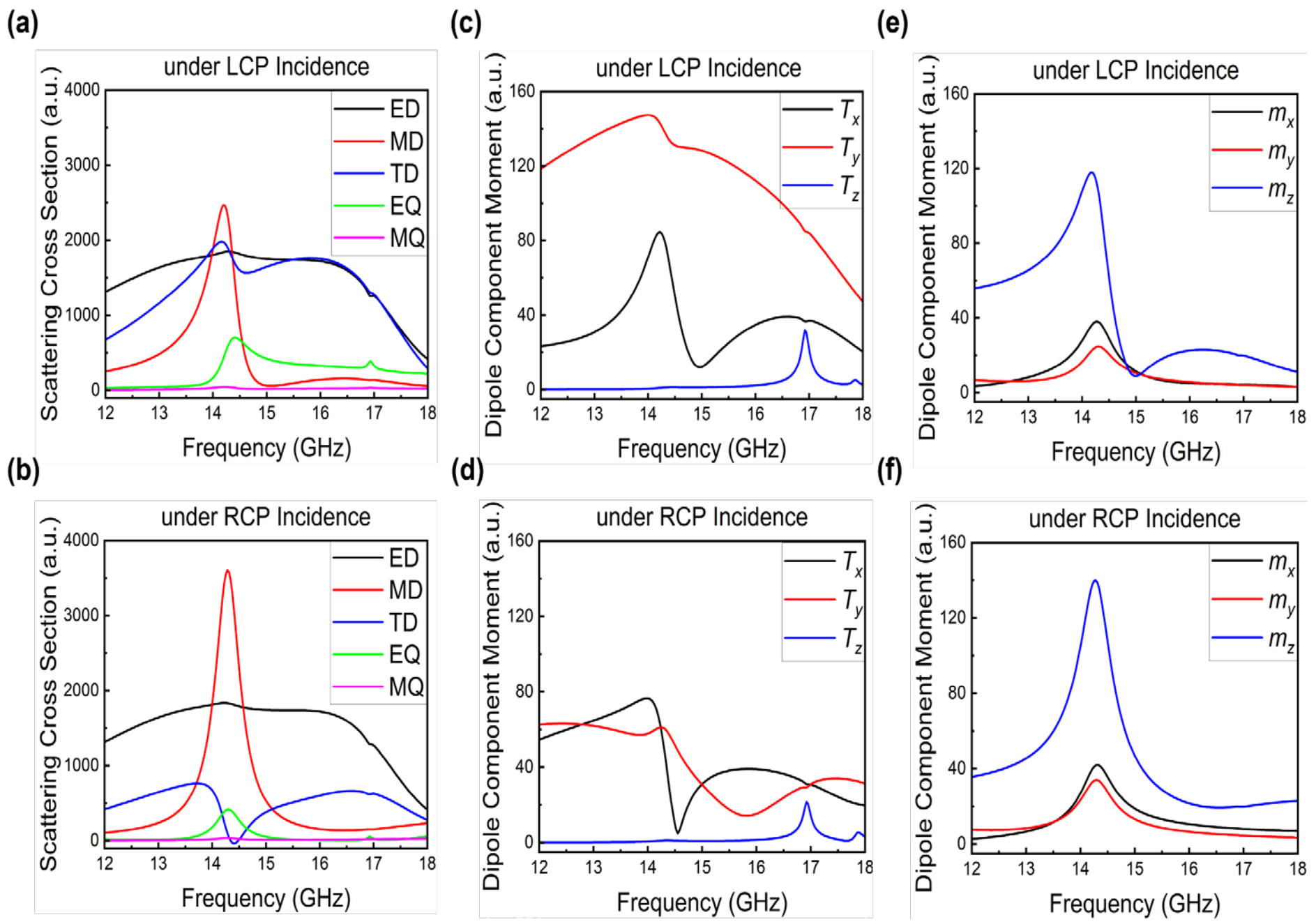

**Fig. 5. The scattering spectra of chiral metasurface under bilayer offset *d* = 5 mm.** (a-b) The calculated scattering cross sections of the lower-order multipoles, including electric dipole (ED), magnetic dipole (MD), toroidal dipole (TD), electric quadrupole (EQ), and magnetic quadrupole (MQ) of the metasurface, under LCP and RCP incidences. (c-d) The amplitude of the TD components under LCP and RCP incidences, including $T_x$, $T_y$ and $T_z$. (e-f) The amplitude of the MD components under LCP and RCP incidences, including $m_x$, $m_y$ and $m_z$.

To have a deeper understanding of the intrinsic giant chiral response exhibited by our designed metasurface, we calculate the corresponding scattering cross sections of the multipoles excited in each unit cell when bilayer offset $d$ = 5 mm (see Supplement 1 Figure S7 for details of achiral situation with $d$ = 0 mm), including electric dipole (ED) moment **P**, magnetic dipole (MD) moment $\mathbf{M}$, toroidal dipole (TD) moment $\mathbf{T}$, electric quadrupole (EQ) moment $\mathbf{Q}_{\alpha\beta}$, and magnetic quadrupole (MQ) moment $\mathbf{M}_{\alpha\beta}$ (see Supplement 1 Note 3 for details of definitions of these dipole moments). The scattering cross sections $C_{sca}$ induced by each dipole moment are defined by the following equations [45-47]

$$\boldsymbol{C}_{sca}^{(\boldsymbol{ED})} = \frac{2\omega^4}{3c^3}|\boldsymbol{P}|^2, \tag{4}$$

$$\boldsymbol{C}_{sca}^{(\boldsymbol{MD})} = \frac{2\omega^4}{3c^3}|\boldsymbol{M}|^2, \tag{5}$$

$$\boldsymbol{C}_{sca}^{(\boldsymbol{TD})} = \frac{4\omega^5}{3c^4}\mathrm{Im}\left(\boldsymbol{P}^* \cdot \boldsymbol{T}\right) + \frac{2\omega^6}{3c^5}|\boldsymbol{T}|^2, \tag{6}$$

$$\boldsymbol{C}_{sca}^{(\boldsymbol{EQ})} = \frac{\omega^6}{5c^5}\sum\left|\boldsymbol{Q}_{\alpha\beta}\right|^2, \tag{7}$$

$$\boldsymbol{C}_{sca}^{(\boldsymbol{MQ})} = \frac{\omega^6}{40c^5}\sum\left|\boldsymbol{M}_{\alpha\beta}\right|^2, \tag{8}$$

where * represents the complex conjugate. According to the above formulas, we calculate the scattering cross sections induced by these multipole moments. As can be seen from the Fig. 5 (a-b), comparing the case of LCP incidence to that of RCP incidence, only the scattering cross sections $C_{sca}$ corresponding to the MD moment and TD moment shows significant differences, thus contributing to the giant chiral response.

Furthermore, around the frequency of 14.3 GHz when the CD of the metasurface reaches the maximum value, the different behaviors of the MD and TD scattering cross sections under LCP and RCP incidences also reach the extreme, especially for the TD moment. It can be seen that around the frequency, under the LCP incidence, the TD scattering cross section reaches the maximum, while under the RCP incidence, the TD scattering cross section approaches zero value. As for the contribution from the EQ moment and MQ moment, it is evident that the scattering spectra induced by the EQ moment exhibit small variations under LCP and RCP incidences. However, these differences are less significant compared to those induced by MD and TD moments and do not play a primary role in generating the strong chirality of the metasurface structure. Additionally, for the MQ moment, its induced scattering spectra under LCP and RCP incidences are almost indistinguishable from the zero line across the entire frequency range. Hence, they are insignificant, and their influence on the chiral response of the metasurface is negligible. Based on the above observations, we can confirm that the giant chiral response with bilayer offset mainly arises from the TD and MD moments.

Therefore, to further reveal the origin of the giant chiral response, we calculate the spatial components of the TD and MD moments [48, 49], $T_x$, $T_y$, $T_z$ and $m_x$, $m_y$, $m_z$, respectively. It is clear from Fig. 5 (c) and 5 (d) that it is not all the three components of the toroidal dipole that dominate this chiral response, but rather it is the toroidal dipole's in-plane components $T_x$, $T_y$, especially $T_y$, that dominate the chiral response. As under LCP and RCP incidences, the intensity of component moment spectrum contributed from $T_x$ and $T_y$ are much larger than that from $T_z$ component, $T_z$ component remains essentially unchanged. In contrast, the out-of-plane components of the magnetic dipole moment, $m_z$, as is shown in Fig. 5 (e) and 5 (f), plays a dominant role in the whole frequency range, while the in-plane components, $m_x$ and $m_y$, show no significant change under LCP and RCP incidences. Based on the above observations, we can conclude that the giant chiral response of the proposed bilayer offset metasurface originates from the out-of-plane magnetic dipole component $m_z$ and the in-plane toroidal dipole components $T_x$, $T_y$ excited around the resonance. In other words, the non-zero offset $d$ creates a chiral near-field distribution that selectively excites in-plane toroidal dipole components ($T_x$ and $T_y$) and out-of-plane magnetic dipole component ($m_z$), thereby inducing a significant CD effect.

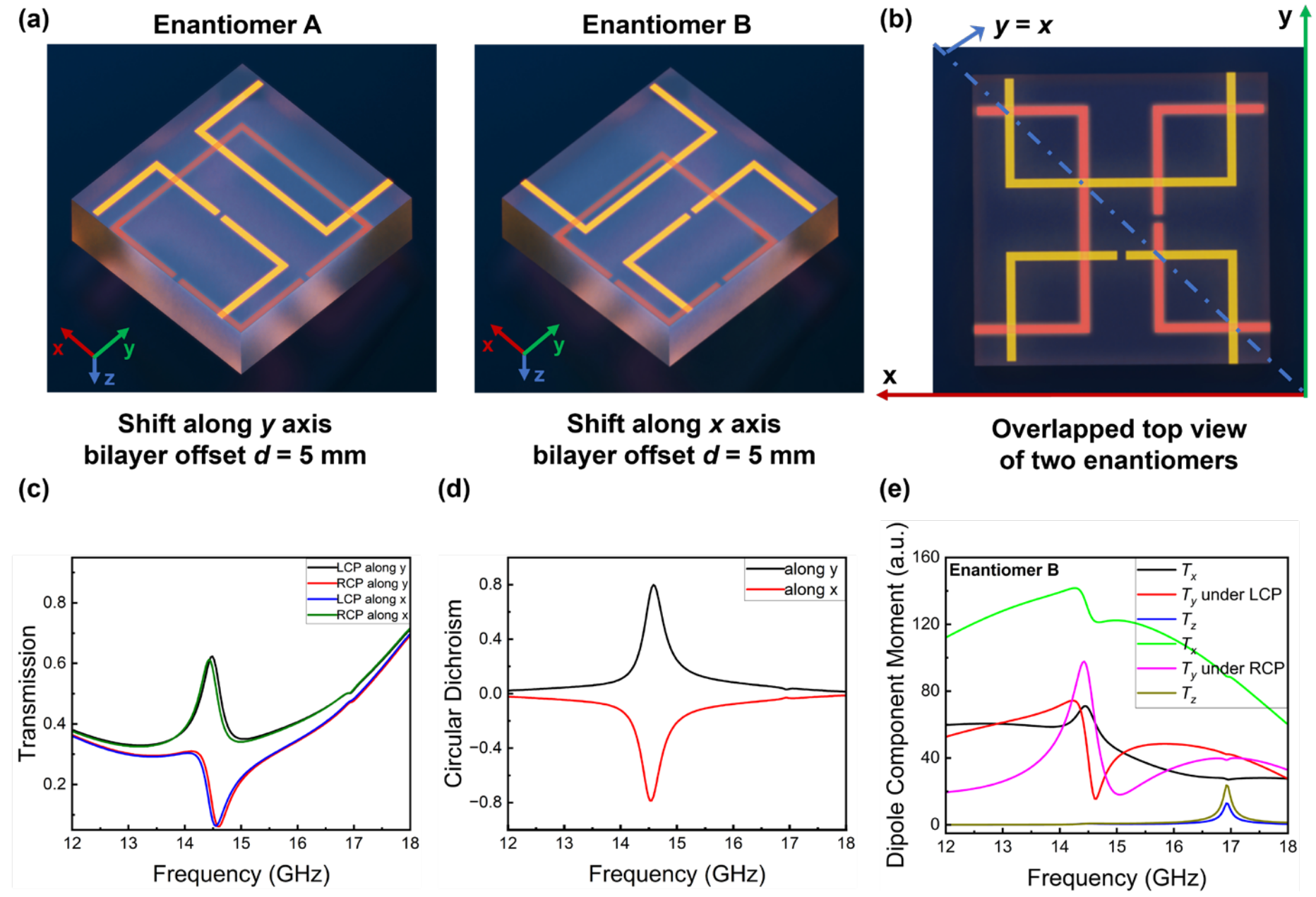

**Fig. 6. Two enantiomers of proposed metasurface with bilayer offset *d* = 5 mm and the opposite circular dichroism (CD) for these two enantiomers.** (a) Comparison and illustration of two enantiomers of the chiral unit cell. Enantiomer A is formed by shifting the top metallic pattern with bilayer offset *d* = 5 mm along the *y* direction (discussed above), while the other one, enantiomer B, is also formed by shifting the top metallic pattern with the same offset *d* value, but along the *x* direction instead. (b) Overlapped top view for the two enantiomers. It can be seen that the two enantiomers can be mapped to each other through a mirror symmetry along the diagonal line (blue dash line). (c) The calculated transmission spectra of enantiomer A and enantiomer B structures under LCP and RCP incidences. (d) The derived CD spectra of two enantiomers. The two enantiomers' CD values are exactly opposite throughout the whole frequency range. (e) The amplitude of the TD components (including $T_x$, $T_y$ and $T_z$) under LCP and RCP incidences for Enantiomer B.

Therefore, it is revealed that the generation of toroidal circular dichroism for the chiral metasurface is owing to the shifting operation of top metallic patterns along the $y$ direction. For the configuration with bilayer offset $d$ = 5 mm showing giant chiral response, we label it as Enantiomer A [50, 51] for convenience of further discussion, showing giant chiral response. Meanwhile, another chiral metasurface can be obtained likewise by changing the shifting direction from $y$ direction to $x$ direction instead for the last step of Fig. 1(c). This configuration, also with bilayer offset $d$ = 5 mm, is labeled as Enantiomer B, as shown in Fig 6 (a). In fact, the two metasurfaces (unit cells) are indeed the mirror image to each other, as shown in Fig. 6 (b), with the mirror axis along the diagonal direction. In order to verify the chiral responses of Enantiomer B, we calculate the total transmission spectra of these two configurations (Enantiomer A and Enantiomer B) under circular incidences separately, as shown in Fig. 6 (c). The results show that the total transmission values of Enantiomer A under LCP incidence conditions and the corresponding Enantiomer B under RCP incidence are equal, and the transmission spectra almost overlap, and the vice versa is also true. However, when we calculate the CD values of this pair of enantiomers accordingly, and we find that they show exactly opposite CDs, as shown in Fig. 6 (d). The opposite CD spectra prove that the proposed metasurface structure can possess a chirality reversal phenomenon if the upper metallic becomes movable, which should be easy to realize with layered structures. Therefore, the results also show that the proposed chiral metasurface has great potential for flexible manipulation of EM waves. As illustrated in Fig. 6 (e), we calculate the three components of the toroidal dipole for Enantiomer B under LCP and RCP wave incidences, including $T_x$, $T_y$, and $T_z$. Our findings reveal that, under LCP wave incidence, the magnitudes of the three components $T_x$, $T_y$, and $T_z$ for Enantiomer B correspond precisely to the results obtained for Enantiomer A under RCP wave incidence. For Enantiomer A whose bilayer offset is along $y$ direction, the difference in $T_y$ dominates, while for Enantiomer B whose bilayer offset is along $x$ direction, the difference in $T_x$ dominates. This observation aligns with the fact that Enantiomer A and Enantiomer B exhibit equal magnitudes but opposite signs in their CD spectra, and confirms in-plane TD components along the bilayer offset direction plays the most significant role in their strong chiral responses.

## 3. Discussion

In summary, we have explored and demonstrated a general approach to achieving pronounced toroidal circular dichroism by designing, manufacturing, and experimentally validating a chiral metasurface composed of purely planar microstructures. This approach utilizes a bilayer offset to naturally break mirror symmetries, leading to a significant chiral response predominantly driven by toroidal dipoles. Our analysis reveals that the giant circular dichroism arises from in-plane components of the toroidal dipoles, with experimental measurements confirming a maximum chiral response of 0.69, closely aligning with simulation results (maximum ±0.80). Notably, the toroidal circular dichroism can be easily reversed by altering the bilayer offset direction. Therefore, this approach of inducing strong toroidal circular dichroism through bilayer offsets, taking advantages of the planar nature and straightforward fabrication, offers new avenues for the development of highly compact and integrated chiral devices. The design mechanism based on bilayer offset should also be readily incorporated into other planar metasurfaces to develop exotic functionalities, such as programmable metasurfaces to enable the development of dynamic chirality.

## Supplemental Material

See Supplement 1 for supporting content.

## Acknowledgments

This work is supported by National Natural Science Foundation of China (No. 12304348),

Guangzhou Municipal Science and Technology Project (No. 2023A03J0003, No. 2024A04J4351), Research on HKUST(GZ) Practices (HKUST(GZ)-ROP2023021), Guangdong Provincial Quantum Science Strategic Initiative (No. GDZX2302005). X. Wu and P. Chen acknowledge support from the Modern Matter Laboratory in HKUST(GZ). P. Chen also acknowledges support from the Quantum Science Center of Guangdong-Hong Kong-Macao Greater Bay Area (Guangdong).

### Conflict of Interest

The authors declare no conflicts of interest.

### Data Availability

Data underlying the results presented in this paper are not publicly available at this time but may be obtained from the corresponding author upon reasonable request.